\documentstyle[11pt]{article}

\def\cl{{\cal L}}
\def\ol#1{\overline{#1}}
\def\lrvec#1{ \stackrel{\leftrightarrow}{#1} }
\def\prt{\partial}
\def\ket#1{|{#1}\rangle}
\def\bra#1{\langle{#1}|}
\def\hmf{ \widehat{m}_F }
\def\tmf{ \widetilde{m}_F }

\def\codt{\cos{\om_s T_s}}
\def\sodt{\sin{\om_s T_s}}
\def\ctodt{\cos{2\om_s T_s}}
\def\stodt{\sin{2\om_s T_s}}
\def\som{\om^\sharp}

\def\al{\alpha}
\def\be{\beta}
\def\ga{\gamma}
\def\de{\delta}
\def\ep{\epsilon}
\def\ze{\zeta}
\def\et{\eta}

\def\ka{\kappa}
\def\la{\lambda}

\def\si{\sigma}

\def\ps{\psi}
\def\om{\omega}
\def\Ga{\Gamma}

\def\Om{\Omega}
\def\mn{{\mu\nu}}

\def\fr#1#2{{{#1} \over {#2}}}
\def\frac#1#2{{\textstyle{{#1}\over {#2}}}}
\def\half{{\textstyle{1\over 2}}}
\def\lsim{\mathrel{\rlap{\lower4pt\hbox{\hskip1pt$\sim$}}
    \raise1pt\hbox{$<$}}}

\def\pr#1{{#1}^\prime}

\def\a{$a_\mu$}
\def\b{$b_\mu$}
\def\c{$c_{\mu\nu}$}
\def\d{$d_{\mu\nu}$}
\def\e{$e_\mu$}
\def\f{$f_\mu$}
\def\g{$g_{\la\mu\nu}$}
\def\H{$H_{\mu\nu}$}

\def\tb{\tilde{b}}
\def\td{\tilde{d}}
\def\tg{\tilde{g}}
\def\tc{\tilde{c}}

\def\som{\om^\sharp}

\def\etal{{\it et al.}}

\newcommand{\beq}{\begin{equation}}
\newcommand{\eeq}{\end{equation}}
\newcommand{\bea}{\begin{eqnarray}}
\newcommand{\eea}{\end{eqnarray}}
\newcommand{\rf}[1]{(\ref{#1})}

\def\secn#1{\vspace{.4cm}\noindent
{\bf {\large\bf #1}}\\[.2cm]}

\def\ci#1{\cite{#1}}
\def\bi#1{\bibitem{#1}}

\begin{document}
\title{Testing Relativity with Orbiting Oscillators
\footnote{Invited talk at
HYPER Symposium: Fundamental Physics and Applications
of Cold Atoms in Space,
Centre National d'Etudes Espatiales,
Paris, France, 4-6 November 2002.}}

\author{
Neil Russell\\
Physics Department\\
Northern Michigan University\\
Marquette, MI 49855, USA}
\date{}
\maketitle

\begin{abstract}
Clock-comparison experiments using a satellite platform
can give Planck-scale sensitivity to many parameters
for Lorentz and CPT violation that are difficult to measure on Earth.
A discussion of the theoretical framework for such
tests is given, with emphasis on comparisons of
output frequencies of atomic clocks and of electromagnetic
cavity oscillators.
\end{abstract}

\secn{Introduction}
Special relativity is an important underlying
foundation for physics.
For almost a century,
experiments with ever-increasing precision
have confirmed the validity of Lorentz symmetry.
As precisions improve,
there remains the possibility of detecting
deviations from special relativity in experiments.
High-precision atomic clocks and cavity oscillators
planned for flight on the International Space Station (ISS)
may be in a unique position to investigate this frontier
in the coming years.

The Standard-Model Extension (SME) is a comprehensive framework
detailing all possible coefficients that quantify Lorentz
violation \ci{cpt01}
and the associated CPT violation \ci{owg}.
The breadth of the SME follows because
all observable signals of Lorentz violation can be described
by effective field theory \ci{sme}.
The extensive literature on the SME in Minkowski spacetime
has been complemented by recent work
to include also gravitational effects \ci{grav}.
A wide variety of experiments and theoretical investigations
have been conducted to place bounds on the coefficients
of the SME.
Areas of physics affected
include neutral mesons and baryogenesis
\ci{hadronexpt,ak,hadronth},
muon properties \ci{muons},
Penning traps \ci{eexpt},
comparisons of hydrogen with antihydrogen \ci{hbar},
and spin-polarized torsion pendula \ci{eexpt2}.
Recent work on neutrinos
\ci{neutrinos,nulong} hints at the possibility of
discovering Lorentz violation in that sector.

In many of these experiments, the sensitivity
has attained the Planck-suppressed levels
at which quantum-gravity effects may be expected on dimensional grounds,
and is therefore of considerable interest.
Despite dozens of experimental investigations,
many of the numerous independent coefficients
that quantify Lorentz violation remain unexplored.
A large number of experiments will be needed to
fully explore the SME parameter space.

The opportunity to place high-precision
clock technology of various types on the
ISS has the potential
to open up several areas of the parameter
space for Lorentz violation.
Clock-comparison experiments will be possible
by measuring the beat frequency between oscillators
of various types, including hydrogen masers,
rubidium and cesium atomic clocks,
and microwave-cavity oscillators.
Clock comparisons provide excellent tests
of Lorentz symmetry \ci{cc}.
A number of such tests have been conducted
in Earth-based laboratories \ci{ccexpt},
and a detailed analysis of prospects for space tests
exists \ci{spaceexpt}.

Cavity oscillators provide access to the photon sector of the SME,
so play a complementary role to the atomic clocks.
There has been much interest in Lorentz violation
in electromagnetism \ci{photonth,klpe}.
The general effects of the SME in electromagnetism are known,
and have yielded an exquisite test of the symmetry,
bounding ten coefficients at the level of parts in 10$^{32}$
\ci{km}.
Various earlier experimental results exist in this sector \ci{photonexpt}.
Cavity oscillators in the microwave and optical regimes
have produced recent bounds on SME coefficients \ci{photon, cavexpt},
and there are plans for experiments of this genre
to orbit on the ISS \ci{sumo}.
This proceedings summarizes aspects of the
SME relevant to the experiments planned on the ISS.
More details can be found in the references.

\secn{Standard-Model Extension (SME)}
In the framework of the SME,
the lagrangian describing a spin-$\half$ Dirac fermion $\ps$ of mass $m$
in the presence of Lorentz violation is
\ci{sme}:
\beq
\cl = \frac{1}{2}i \ol{\ps} \Ga_\nu \lrvec{\prt^\nu} \ps
   - \ol{\ps} M \ps
\quad ,
\label{lagr}
\eeq
where
\bea
M &:=& m + a_\mu \ga^\mu + b_\mu \ga_5 \ga^\mu
   + \half H_\mn \si^\mn
\quad \mbox{and}
\label{M} \\
\Ga_\nu &:=& \ga_\nu + c_\mn \ga^\mu + d_\mn \ga_5 \ga^\mu
+ e_\nu + i f_\nu \ga_5 + \half g_{\la \mu \nu} \si^{\la \mu}
\quad .
\label{Gam}
\eea

The conventional Lorentz-preserving case is recovered
from just the first term in each of $M$ and $\Ga_{\nu}$ above.
The additional terms in equations $M$ and $\Ga_{\nu}$ contain
conventional Dirac matrices
$\left\{1, \ga_5, \ga^\mu, \ga_5\ga^\mu, \si^{\mu\nu} \right\}$.
They also contain parameters
\a, \b, \c, \d, \e, \f, \g, and \H,
that imply Lorentz violation in equation \rf{lagr}.
Various mechanisms giving rise to these parameters are possible.
They could for example arise as expectation values of Lorentz tensors
in a fundamental theory with spontaneous Lorentz breaking \ci{akss}.
The parameters in $M$ have dimensions of mass,
and those in $\Ga_{\nu}$ are dimensionless;
\c\ and \d\ are traceless,
while \H\ is antisymmetric
and \g\ is antisymmetric in its first two indices.

The Lorentz-violation parameters in the lagrangian
can be thought of as fixed geometrical background objects
in spacetime.
An experiment that rotates in space could in principle detect
time-dependent projections of these geometric quantities.
Similarly, two identical experiments with differing relative
velocities could discern boost-dependent effects.
Thus Lorentz violation is seen through comparisons of identical
experiments with differing rotations and boosts,
or through time dependence in a single experiment with nonzero acceleration.
Lorentz symmetry is violated under these `particle transformations'
of entire experimental configurations.
In contrast,
experimenters observing one experimental system from
different boosted or rotated inertial reference frames will
find that the components of the parameters
\a, \b, \c, \d, \e, \f, \g, and \H\
transform like conventional tensors under
Lorentz transformations of the coordinates.
Thus, the SME preserves every aspect of
conventional observer Lorentz symmetry.

The Lorentz-violation parameters are known to be minuscule,
and so this formalism is well suited to treatment within
perturbation theory.
The theory has been studied in various contexts \ci{bek,cm,ark}.
It is possible to pass to a hamiltonian formalism
and, with appropriate assumptions for the effective fermion $\psi$,
to find the energy-level corrections for atoms within
an atomic clock.

\secn{General Clock-comparison Experiments}
An atomic clock operates by producing a stable output angular frequency
based on an atomic energy-level transition.
In many cases this frequency depends
on a magnetic field that forms the quantization axis.
Thus,
if the third coordinate is defined to lie along this quantization axis,
then the output frequency is $f(B_3)$.
Stability of the clock is increased by operating the clock
near a field-independent point and keeping the field $B_3$
as constant as possible.

In the presence of Lorentz violation, the SME
provides a general framework that shifts the clock frequency,
giving
\beq
\om = f(B_3) + \de \om \ .
\label{om}
\eeq
The quantity $\de\om$ contains all the contributions from Lorentz-violating
terms in the SME lagrangian.

This small correction can contain terms that are orientation dependent,
such as for example the dot product of $\vec{B}$ and the spatial part of
$b^{\mu}$.
It can also depend on the boost velocity of the clock relative to the
inertial reference frame in which the four-vector $b^{\mu}$ is expressed.
Even though the clock is usually stationary in the laboratory,
the laboratory itself is a moving reference frame
relative to the inertial reference frame.
The exact form of $\de\om$ can be complicated and involves
corrections for all the Lorentz-violating added terms in the SME lagrangian
and requires a detailed knowledge of the motion of the clock laboratory.

In many realistic cases, the function $f$ can be inverted for values of $B_3$
ranging over those encountered experimentally.
We therefore consider cases where $f^{-1}$ exists,
and note that cases where the clock operates at a field-independent point
can be handled by alternative methods.
If the form of $f$ is known within the framework of conventional
Lorentz-preserving physics,
then for a clock running with frequency $\om$ in the presence of Lorentz violation
the inverse $f^{-1}(\om)$ gives an effective magnetic field
differing slightly from the actual magnetic field.
If there is no Lorentz violation,
this effective magnetic field is the actual $B_3$.

To search for evidence of Lorentz violation in practical terms
means that the clock frequency has to be compared to
a standard, which is essentially another atomic clock.
Thus the comparison is made between two clocks, with frequencies
$\om_A$ and $\om_B$, both of which are sensitive to Lorentz violation.
One way to seek violations would be to monitor
the frequency difference $\om_A - \om_B$.
Any time dependence would indicate Lorentz violation.
In conventional physics,
this difference is non-zero and equals
$f_A(B_3)-f_B(B_3)$.
If the functions $f_A$ and $f_B$ have matching slopes
then the difference would be constant
even with variations in $B_3$.
This matching of slopes is not in general possible
for a given pair of atomic clocks.

A preferable measure of Lorentz violation that
circumvents the difficulties with magnetic-field
dependence is the modified frequency difference
\begin{equation}
\om^\sharp := \om_A - f_A \circ f_B^{-1}(\om_B)
\quad .
\label{omsharpdef}
\end{equation}
Using simple differentiation assumptions on the functions $f_A$ and $f_B$,
it follows \ci{spaceexpt} that $\om^{\sharp}$ is independent of $B_3$,
equalling
\begin{equation}
\om^{\sharp} = \de \om_A - v \de \om_ B
\quad ,
\label{omsharp}
\end{equation}
where the quantity $v$ is the dimensionless constant ratio
of the two frequency gradients evaluated at zero field:
\begin{equation}
v = \left( \fr {df_A}{dB_3}\bigg/\fr{df_B}{dB_3}\right)
    \bigg\vert_{B_3=0}
\quad .
\end{equation}
Note also that in the absence of Lorentz violation,
$\om^{\sharp}$ vanishes.

To utilize equation (\ref{omsharp}), several options are possible.
If the functions $f_A$ and $f_B$ are known in detail,
then one way to proceed is to record the values of $\om_A$ and $\om_B$
at each instant, and then to combine them using this equation.
This gives an experimental value of $\som$
that can be compared with the theoretical calculation.
This method requires a detailed knowledge of the functions
$f_A$ and $f_B$, which may not be possible in practise.

An alternative method to experimentally determine $\som$
is to use feedback control to keep the frequency $\om_B$ of clock B
fixed relative to itself.
Then in equation (\ref{omsharpdef}),
$f_A[f_B^{-1}(\om_B)]$ is constant,
making $\som=\om_A +$~constant.
The constant value that arises even in the absence of
Lorentz violation is irrelevant, since the experimental
procedure requires monitoring only the variations in $\som$.
This method could be useful in situations where a detailed knowledge
of $f_A$ and $f_B$ is not known but the two clocks are
in the same magnetic field $B_3$.
This may offer advantages for potential clock-comparison experiments
on the ISS, where the magnetic field
is likely to fluctuate.
The signal to be monitored would thus be
\begin{equation}
\om_A =\som - \mbox{constant}=\de\om_A-v\de\om_B - \mbox{constant}
\quad .
\end{equation}

\secn{Applying the SME in the Laboratory Frame}
The atoms comprising an atomic clock
are a complex system of protons, neutrons, and electrons.
To calculate the effect of the SME on such systems,
a variety of simplifying assumptions are made to model the
system with a single wave function $\psi$.
The hamiltonian for this system can be split into two portions:
a conventional part describing the atom within the chosen model,
and a perturbative Lorentz-violating part $\pr{h}$
arising from the SME.
It can be expressed as a sum of perturbative hamiltonians for
each proton, electron, and neutron (indexed by $w$) in the atom:
\beq
\pr{h}=\sum_w\sum_{N=1}^{N_w} \de {h}_{w,N}
\quad .
\label{hprime}
\eeq
In this expression, the atom or ion $W$ has $N_w$ particles
of type $w$,
and $\de {h}_{w,N}$ is the Lorentz-violating correction
for the $N$th particle of type $w$.
Since each of the three particle species in the atom has
a set of Lorentz-violation parameters,
a superscript $w$ must be placed on each of the parameters
\a, \b, \c, \d, \e, \f, \g, and \H.

The symmetry-breaking energy-level shifts are calculated
by finding the expectation value of the perturbative
hamiltonian $\pr{h}$ in the desired unperturbed state of the atoms.
In most cases,
the quantization axis is defined by a magnetic field,
and the total angular momentum $\vec F$ of the atom or ion
and its projection along the quantization axis
are conserved to a good approximation.
Thus quantum states for the atomic-clock atoms
can be labelled by the corresponding quantum numbers $\ket{F,m_F}$.
We define the third coordinate of the laboratory reference frame
to be this quantization axis.

In the laboratory frame, the energy-level shift for state
$\ket{F,m_F}$ is
\bea
\de E(F,m_F) &=&
\bra{F,m_F} \pr{h} \ket{F,m_F}
\nonumber \\
&=& \hmf \sum_w
(\be_w\tb_3^w + \de_w\td_3^w + \ka_w\tg_d^w)
+ \tmf \sum_w (\ga_w\tc_q^w + \la_w\tg_q^w) \ .
\label{AtomicShift}
\eea
In this expression,
$\hmf$ and $\tmf$ are particular ratios
of Clebsch-Gordan coefficients.
The quantities
$\be_w$, $\de_w$, $\ka_w$, $\ga_w$, and $\la_w$
are expectation values of combinations
of spin and momentum operators in the extremal states
$\ket{F,m_F=F}$.
They can not in general be calculated exactly since
a detailed description of the nuclear forces is not known.
For further details of these quantities, see reference
\ci{cc}.

In equation (\ref{AtomicShift}) the quantities with tildes
are specific combinations of Lorentz-violation parameters,
and importantly, are the only possible parameter combinations
to which clock-comparison experiments are sensitive.
For the case of $\tb_3^w$, the combination is
\beq
\tb_3^w := b_3^w -m_w d_{30}^w
 + m_w g_{120}^w -H_{12}^w
 \quad.
\eeq
The other tilde quantities can be found in reference \ci{cc}.

Having found the shifts of the energy levels $\ket{F,m_F}$,
the effect of the SME on the frequency
corresponding to the transition $(F,m_F) \rightarrow (\pr{F},\pr{m}_F)$
is found from the difference
\beq \de\om = \de E(F,m_F)- \de E(\pr{F},\pr{m}_F)
\quad .
\label{FrequencyComparison}
\eeq
This is the $\de\om$ appearing in equation (\ref{om}).

\secn{The Standard Inertial Reference Frame}
The SME indicates that Lorentz violation can occur in nature
through the existence of a variety of background
observer Lorentz tensors.
The objective of experimental tests of Lorentz symmetry
is to measure these tensor components,
or to place bounds on them if experiments are not able to resolve
them.
The inertial reference frame in which the components of these tensors
are measured needs to be standardized to allow different experiments
to compare independent measurements of the same quantities.

By convention, the inertial reference frame used
to present results for bounds on SME quantities
has origin at the center of the Sun.
The axes are labelled $X$, $Y$, and $Z$,
with $Z$ axis parallel to the axis of the Earth in the northerly
orientation.
The $X$ axis points towards the vernal equinox on the celestial sphere,
and the $Y$ axis completes the right-handed system.
This frame is close to inertial over periods of thousands of years,
as opposed to any Earth-based frame in which the inertial approximation breaks
down after a week or two.

The time variable in this frame is denoted by $T$,
and is measured by a clock considered to be at the center of the Sun
with $T=0$ taken to be at the vernal equinox in the year 2000.

We define the laboratory coordinate system to
have third coordinate along the quantization axis.
The laboratory frame $(x_1, x_2, x_3)$ is not in general inertial,
and quantities measured in the laboratory frame
must be transformed into the standard inertial reference frame.
The specifics of the transformation to the inertial reference frame
will depend on the experiment.
Clock-comparison experiments have been done in Earth-based laboratories,
and others are planned for satellites encircling the Earth.
We focus here on the latter.

For our purposes,
the motion of a satellite-based experiment can be considered to
be a superposition of two circular motions,
where one is the motion of the Earth around the Sun,
and the other is the motion of the satellite around the Earth.
The Earth moves on a circle in a plane tilted
at angle $\et \approx 23^{\circ}$ to the equatorial plane,
passing through the positive $X$ axis at the vernal equinox.
The orbit of the satellite can be specified by giving
the inclination angle $\ze$ to the inertial $Z$ axis,
and the right ascension $\al$ of the ascending node of the orbit,
where the satellite cuts the equatorial plane in the northward direction.

There are various possibilities for the orientation of the
laboratory reference frame within the satellite.
For definiteness,
we define the quantization axis $x_3$ to be directed along the
velocity vector of the satellite relative to the Earth,
and the $x_1$ axis to be directed radially towards the center of the Earth.

With these definitions, the laboratory-frame observable quantities
in clock-comparison experiments can be expressed in
terms of the corresponding inertial-frame quantities.
Since the motion is a composition of two circular motions,
the expressions are long,
even with simplifying assumptions such as zero eccentricity for the orbits.
The general form for the expression $\tb_3$ is:

\begin{footnotesize}
\bea \tb_3 &=&
\codt \Big\{ \Big[
 -\tb_X \sin\al\cos\ze + \tb_Y\cos\al\cos\ze + \tb_Z\sin\ze
 \Big]
+\be_\oplus[\mbox{seasonal terms}\ldots] \Big\}
\nonumber \\
&+& \sodt \Big\{ \Big[
 -\tb_X \cos\al - \tb_Y \sin\al
 \Big]
 +\be_\oplus[\mbox{seasonal terms}\ldots] \Big\}
 \nonumber \\
&+& \ctodt \Big\{
 \be_s[\mbox{constant terms}\ldots] \Big\}
+ \stodt \Big\{
 \be_s [\mbox{constant terms}\ldots] \Big\}
 \nonumber \\
&+&
 \be_s[\mbox{constant terms}\ldots ] \ .
\label{Explicitb3}
\eea
\end{footnotesize}
Oscillations involving the satellite orbital frequency
appear with single and double frequencies $\om_s$, and $2\om_s$.
The much slower orbital frequency $\Om_\oplus$ of the Earth
as it moves around the Sun
appears in the seasonal terms as indicated.
The suppressions due to the speed $\be_\oplus \approx 10^{-4}$ of the Earth
and $\be_s \approx 10^{-5}$ for the ISS are also shown explicitly.
For further details of $\tb_3$,
see \cite{spaceexpt},
and see \cite{cc} for details concerning the nonrelativistic limit.

The energy shift expressed in equation (\ref{AtomicShift})
also depends on the specific atoms in the clock
and the transition being used for the clock.
Species planned for flight on the ISS
include rubidium 87, cesium 133, and hydrogen.
Further details are available in \ci{spaceexpt}.

\secn{The Photon Sector}
Another highly-stable clock is the cavity oscillator,
in which a resonant frequency of an electromagnetic
oscillation is excited.
There is interest in such oscillators for use
on the ISS
\ci{sumo}.
The SME provides a complete and unified framework
detailing all possible Lorentz violations
in the photon sector\ci{km}.
The photon-sector lagrangian can be expressed as
\bea
\cl&=&\half[(1+\tilde\ka_{\rm tr})\vec E^2
-(1-\tilde\ka_{\rm tr})\vec B^2]
\label{lagkappa2}\\
&&
+\half \vec E\cdot(\tilde\ka_{e+}+\tilde\ka_{e-})\cdot\vec E
-\half\vec B\cdot(\tilde\ka_{e+}-\tilde\ka_{e-})\cdot\vec B
+\vec E\cdot(\tilde\ka_{o+}+\tilde\ka_{o-})\cdot\vec B .
\nonumber
\eea
Here,
$\tilde\ka_{\rm tr}$ is a single number,
$\tilde\ka_{e+}$, $\tilde\ka_{e-}$, $\tilde\ka_{o-}$
are traceless symmetric $3\times 3$ matrices,
and $\tilde\ka_{o+}$ is an antisymmetric $3\times 3$ matrix,
giving a total of 19 independent coefficients for Lorentz violation
in the photon sector.
The $\tilde \ka$ quantities are tensor-like geometric quantities that
can in principle be detected with electromagnetic experiments.

This lagrangian leads to modified Maxwell equations
and hence to a modified wave equation for light.
Among the effects is a rotation of the polarization vector
that is frequency dependent.
To observe such a rotation,
the propagation distance should be maximized.
Using polarization data for light from 16 distant cosmological
sources,
a bound \ci{km} has been
placed on 10 linear combinations of the $\tilde \ka$ coefficients,
denoted by $k^a$ for $a=1, \ldots, 10$.
The result is
\beq
|k^a|<2 \times 10^{-32} \ .
\label{polbound}
\eeq
These bounds are among the tightest bounds in the photon sector,
but apply only to 10 linear combinations of coefficients.
Of the 9 remaining independent coefficients,
seven have been bounded
in experiments with electromagnetic cavity oscillators
at the level of parts in $10^{15}$ \ci{photon}.

For resonant cavities,
the SME lagrangian (\ref{lagkappa2})
leads to a fractional resonant-frequency shift
$\de \nu / \nu$
that depends on the
geometrical relationship between the
axis of the cylindrical cavity
and the SME background tensor-like fields.
For optical frequencies, the general form of the
relative frequency shift is
\beq
\fr{\de\nu}{\nu}
= -\fr1 {2|\vec E_0|^2}
\bigl[\vec E_0^*\cdot(\ka_{DE})_{\rm lab}\cdot\vec E_0/\ep
-(\hat N \times \vec E_0^*)
\cdot(\ka_{HB})_{\rm lab}\cdot(\hat N \times \vec E_0)\bigr] ,
\label{dnuopt}
\eeq
where $\hat N$ is the axis of the cylindrical cavity,
$\vec{E}_0$ is the electric field vector,
and $\ka_{DE},$ $\ka_{HB}$ are specific linear combinations
of the $\tilde{\ka}$ matrices in the laboratory reference frame.

Since the laboratory is noninertial,
being either on the surface of the Earth or on
a satellite orbiting the Earth,
equation (\ref{dnuopt}) must be expressed in terms of
the standard inertial reference frame as discussed
for atomic clocks above.
If the laboratory is Earth-based,
the oscillator output contains sinusoidal
variations with frequencies $\om_\oplus$ and $2 \om_\oplus$
as well as slower seasonal variations due to the tilt
of the Earth relative to the Sun.
Similar single- and double-frequency effects
can be expected for oscillators on a satellite.
Several bounds on these coefficients have been
obtained in Earth-based experiments
using optical and microwave frequencies \ci{photon, cavexpt}.

\secn{Discussion}
No evidence exists for Lorentz violation at present,
although an interesting possibility relating to the
apparent mass of neutrinos may change this \ci{neutrinos,nulong}.
The SME provides the full parameter space for
testing Lorentz symmetry.
While parts of this space have been investigated
with experiments from all corners of the physics globe,
there are still some regions that are inaccessible.
Experiments on the ISS
will open up a path to several outlying regions of this world.
Atomic clocks \ci{spaceexpt}
and microwave cavities \ci{km, photon, cavexpt}
aboard the ISS
may give a new perspective
on whether nature is Lorentz symmetric.


\begin{thebibliography}{xx}

\bi{cpt01}
An overview of Lorentz and CPT Violation
is given, for example, in the volume
V.A.\ Kosteleck\'y, ed.,
\it CPT and Lorentz Symmetry II, \rm
World Scientific, Singapore, 2002.

\bibitem{owg}
O.W.\ Greenberg,
Phys.\ Rev.\ Lett.\ {\bf 89}, 231602 (2002);
Phys.\ Lett.\ B {\bf 567}, 179 (2003).

\bi{sme}
V.A.\ Kosteleck\'y and R.\ Potting,
Phys.\ Rev.\ D {\bf 51}, 3923 (1995);
D.\ Colladay and V.A.\ Kosteleck\'y,
Phys.\ Rev.\ D {\bf 55} 6760 (1997);
Phys.\ Rev.\ D {\bf 58} 116002 (1998).

\bi{grav}
V.A.\ Kosteleck\'y,
Phys.\ Rev.\ D {\bf 69}, 105009 (2004).

\bi{hadronexpt}
KTeV Collaboration,
H.\ Nguyen, in Ref.\ \cite{cpt01};
OPAL Collaboration,
R.\ Ackerstaff \etal,
Z.\ Phys.\ C {\bf 76}, 401 (1997);
DELPHI Collaboration,
M.\ Feindt \etal,
preprint DELPHI 97-98 CONF 80 (1997);
BELLE Collaboration,
K.\ Abe \etal,
Phys.\ Rev.\ Lett.\ {\bf 86}, 3228 (2001);
BaBar Collaboration,
B.\ Aubert
\etal,
hep-ex/0303043;
FOCUS Collaboration,
J.M.\ Link \etal,
Phys.\ Lett.\ B {\bf 556}, 7 (2003).

\bi{ak}
V.A.\ Kosteleck\'y,
Phys.\ Rev.\ Lett.\ {\bf 80}, 1818 (1998);
Phys.\ Rev.\ D {\bf 61}, 016002 (2000);
Phys.\ Rev.\ D {\bf 64}, 076001 (2001).

\bi{hadronth}
D.\ Colladay and V.A.\ Kosteleck\'y,
Phys.\ Lett.\ B {\bf 344}, 259 (1995);
Phys.\ Rev.\ D {\bf 52}, 6224 (1995);
Phys.\ Lett.\ B {\bf 511}, 209 (2001);
V.A.\ Kosteleck\'y and R.\ Van Kooten,
Phys.\ Rev.\ D {\bf 54}, 5585 (1996);
O.\ Bertolami \etal,
Phys.\ Lett.\ B {\bf 395}, 178 (1997);
N.\ Isgur \etal,
Phys.\ Lett.\ B {\bf 515}, 333 (2001).

\bi{muons}
V.W.\ Hughes \etal,
Phys.\ Rev.\ Lett.\ {\bf 87}, 111804 (2001);
R.\ Bluhm \etal,
Phys.\ Rev.\ Lett.\ {\bf 84}, 1098 (2000).

\bi{eexpt}
H.\ Dehmelt \etal,
Phys.\ Rev.\ Lett.\ {\bf 83}, 4694 (1999);
R.\ Mittleman \etal,
Phys.\ Rev.\ Lett.\ {\bf 83}, 2116 (1999);
G.\ Gabrielse \etal,
Phys.\ Rev.\ Lett.\ {\bf 82}, 3198 (1999);
R.\ Bluhm \etal,
Phys.\ Rev.\ Lett.\ {\bf 79}, 1432 (1997);
Phys.\ Rev.\ D {\bf 57}, 3932 (1998).

\bi{hbar}
R.\ Bluhm \etal,
Phys.\ Rev.\ Lett.\ {\bf 82}, 2254 (1999);

\bi{eexpt2}
B.\ Heckel,
in Ref.\ \cite{cpt01};
L.-S.\ Hou, W.-T.\ Ni, and Y.-C.M.\ Li,
Phys.\ Rev.\ Lett.\ {\bf 90}, 201101 (2003);
R.\ Bluhm and V.A.\ Kosteleck\'y,
Phys.\ Rev.\ Lett.\ {\bf 84}, 1381 (2000).

\bi{neutrinos}
S.\ Coleman and S.L.\ Glashow,
Phys.\ Rev.\ D {\bf 59}, 116008 (1999);
V.\ Barger, S.\ Pakvasa, T.J.\ Weiler, and K.\ Whisnant,
Phys.\ Rev.\ Lett.\ {\bf 85}, 5055 (2000);
J.N.\ Bahcall, V.\ Barger, and D.\ Marfatia,
Phys.\ Lett.\ B {\bf 534}, 114 (2002).

\bi{nulong}
V.A.\ Kosteleck\'y and M.\ Mewes, hep-ph/0308300;
Phys.\ Rev.\ D {\bf 69}, 016005 (2004).

\bi{cc}
V.A.\ Kosteleck\'y and C.D.\ Lane,
Phys.\ Rev.\ D {\bf 60}, 116010 (1999);
J.\ Math.\ Phys.\ {\bf 40}, 6245 (1999).

\bi{ccexpt}
L.R.\ Hunter \etal,
in
V.A.\ Kosteleck\'y, ed.,
\it CPT and Lorentz Symmetry, \rm
World Scientific, Singapore, 1999;
D.\ Bear \etal,
Phys.\ Rev.\ Lett.\ {\bf 85}, 5038 (2000);
D.F.\ Phillips \etal,
Phys.\ Rev.\ D {\bf 63}, 111101 (2001);
M.A.\ Humphrey \etal,
Phys.\ Rev.\ A {\bf 68}, 063807 (2004);
Phys.\ Rev.\ A {\bf 62}, 063405 (2000);
F.\ Can\`e \etal,
physics/0309070.

\bi{spaceexpt}
R.\ Bluhm \etal,
Phys.\ Rev.\ Lett.\ {\bf 88}, 090801 (2002);
Phys.\ Rev.\ D {\bf 68}, 125008 (2003).

\bi{photonth}
R.\ Jackiw and V.A.\ Kosteleck\'y,
Phys.\ Rev.\ Lett.\ {\bf 82}, 3572 (1999);
C.\ Adam and F.R.\ Klinkhamer,
Nucl.\ Phys.\ B {\bf 657}, 214 (2003);
H.\ M\"uller, C.\ Braxmaier, S.\ Herrmann,
A.\ Peters, and C.\ L\"ammerzahl,
Phys. Rev. D {\bf 67}, 056006 (2003);
T.\ Jacobson, S.\ Liberati, and D.\ Mattingly,
Phys.\ Rev.\ D {\bf 67}, 124011 (2003);
V.A.\ Kosteleck\'y, C.D.\ Lane, and A.G.M.\ Pickering,
Phys.\ Rev.\ D {\bf 65}, 056006 (2002);
V.A.\ Kosteleck\'y and A.G.M.\ Pickering,
Phys.\ Rev.\ Lett.\ {\bf 91}, 031801 (2003);
R.\ Lehnert,
Phys.\ Rev.\ D {\bf 68}, 085003 (2003);
G.M.\ Shore,
Contemp.\ Phys.\ {\bf 44}, 503 {2003};
B.\ Altschul,
Phys.\ Rev.\ D {\bf 69}, 125009 (2004).

\bi{klpe}
V.A.\ Kosteleck\'y, M.\ Perry, and R.\ Lehnert,
Phys.\ Rev.\ D {\bf 68}, 123511 (2003).

\bi{km}
V.A.\ Kosteleck\'y and M.\ Mewes,
Phys.\ Rev.\ Lett.\ {\bf 87}, 251304 (2001);
Phys.\ Rev.\ D {\bf 66}, 056005 (2002).

\bi{photonexpt}
S.M.\ Carroll, G.B.\ Field, and R.\ Jackiw,
Phys. Rev. D {\bf 41}, 1231 (1990);
M.P.\ Haugan and T.F.\ Kauffmann,
Phys. Rev. D {\bf 52}, 3168 (1995).

\bi{photon}
H.\  M\"uller, S.\ Herrmann, C.\ Braxmaier, S.\ Schiller, and A.\ Peters,
Phys.\ Rev.\ Lett.\ {\bf 91}, 020401 (2003).

\bi{cavexpt}
J.\ Lipa, J.\ Nissen, S.\ Wang, D.\ Stricker, and D.\ Avaloff,
Phys.\ Rev.\ Lett.\ {\bf 90}, 060403 (2003);
H.\  M\"uller, S.\ Herrmann, A.\ Saenz, A.\ Peters, and C.\ L\"ammerzahl,
Phys. Rev. D {\bf 68}, 116006 (2003);
P.\ Wolf, M.\ Tobar, S.\ Bize, A.\ Clairon, A.\ Luiten, and G.\ Santarelli,
gr-qc/0401017.

\bi{sumo}
S.\ Buchman {\it et al.}, Adv.\ Space Res.\ {\bf 25}, 1251 (2000);
J.\ Nissen {\it et al.}, in Ref.\ \cite{cpt01}.

\bi{akss}
V.A.\ Kosteleck\'y and S.\ Samuel,
Phys.\ Rev.\ Lett.\ {\bf 63}, 224 (1989);
Phys.\ Rev.\ D {\bf 40}, 1886 (1989);
V.A.\ Kosteleck\'y and R.\ Potting,
Nucl.\ Phys.\ B {\bf 359}, 545 (1991).

\bibitem{bek}
M.S.\ Berger and V.A.\ Kosteleck\'y,
Phys.\ Rev.\ D {\bf 65}, 091701(R) (2002).

\bibitem{cm}
D.\ Colladay and P.\ McDonald,
J.\ Math.\ Phys.\ {\bf 43}, 3554 (2002);
hep-th/0312058.

\bibitem{ark}
V.A.\ Kosteleck\'y and A.\ Roberts,
Phys.\ Rev.\ D {\bf 63}, 096002 (2001).

\end{thebibliography}
\end{document}